\documentstyle[preprint,aps]{revtex}

\tightenlines

\begin{document}
\draft
\title{A Model for Persistent Levy Motion }
\author{A. V. Chechkin and V. Yu. Gonchar}
\address{Institute for Theoretical Physics\\
National Science Center ``Kharkov Institute of Physics and Technology`` \\
Akademicheskaya St.1, Kharkov 310108, Ukraine }
\maketitle

\begin{abstract}
We propose the model, which allows us to approximate fractional Levy noise
and fractional Levy motion. Our model is based (i) on the Gnedenko limit
theorem for an attraction basin of stable probability law, and (ii) on
regarding fractional noise as the result of fractional
integration/differentiation of a white Levy noise. We investigate self -
affine properties of the approximation and conclude that it is suitable for
modeling persistent Levy motion with the Levy index between 1 and 2.
\end{abstract}


\section{Introduction.}

By Levy motions, or Levy processes, we designate a class of random
functions, which are a natural generalization of the Brownian motion, and
whose increments are stationary, statistically self-affine and stably
distributed in the sense of P. Levy \cite{Lev37}. Two important subclasses
are (i) the stable processes, or the ordinary Levy motions, which generalize
the ordinary Brownian motion, or the Wiener process, and whose increments
are independent, and (ii) the fractional Levy motions, which generalize the
fractional Brownian motions and have an infinite span of interdependence.

The theory of processes with independent increments was developed beginning
from the Bachelier's paper \cite{Bac00} concerning Brownian motion. However,
the rigorous construction of this process and studies of properties of its
trajectories were undertaken by Wiener \cite{Wie23}. The modern presentation
of the general theory of processes with independent increments is contained
in \cite{Sko64,Sam94}. The theory of stable processes with independent
increments has begun its history from the already cited work \cite{Lev37}
and, later on, was developed by other prominent mathematicians. In
particular, the properties of extrema of stable symmetric processes were
studied in Ref. \cite{Dar56}. The local properties of stable processes and
geometric properties of their trajectories were considered in Ref. \cite
{Blu60}.

The basis of the general theory of random processes with stationary
self-affine increments was laid by Kolmogorov \cite{Kol40}. Fractional
Brownian motions were introduced by Mandelbrot and van Ness as a
(relatively) simple family of random functions ''that could be in some way
be expected to be ''typical'' of what happens in the absence of asymptotic
independence ''\cite{Man68}. In this paper the possibility was also pointed
out of constructing ''fractional Levy-stable random functions'' in a way
analogous to that of constructing fractional Brownian motion. Such stable
generalizations of fractional Brownian motion were introduced in \cite
{Taq83,Mae83}.

The Levy random processes play an important role in different areas of
application for at least two reasons.

The first one is that the Levy motion can be considered as a generalization
of the Brownian motion. Indeed, the mathematical foundation of the
generalization are remarkable properties of stable probability laws. From
the limit theorems point of view, the stable distributions are a
generalization of widely used Gaussian distribution. Namely, stable
distributions are the limit ones for the distributions of (properly
normalized) sums of independent identically distributed (i.i.d.) random
variables \cite{Gne49}. Therefore, these distributions (like the Gaussian
one) occur, when the evolution of a physical system or the result of an
experiment are determined by the sum of a large number of identical
independent random factors. An important distinction of stable probability
densities is the power law tails decreasing as $\left| x\right| ^{-1-\alpha
},\quad x\rightarrow \infty $ , $\alpha $ is the Levy index, $0<\alpha <2$.
Hence, the distribution moments of the order $\alpha $ diverge. In
particular, stably distributed variables possess a non-finite variance.

The second reason for ubiquity of the Levy motions is their remarkable
property of scale - invariance. From this point of view the Levy motions
(like the Brownian ones) belong to the so - called fractal random processes.
Indeed, the objects in nature rarely exhibit exact self - similarity ( like
the von Koch curve), or self - affinity. On the contrary, these properties
have to be understood in a probabilistic sense \cite{Man82,Vos85}. The
random fractals are believed to be widely spread in nature. A coastline is a
simple example of statistically self - similar object, whereas the trace of
the Brownian motion is statistically self - affine. Several numerical
algorithms were developed in order to simulate fractional Brownian motion 
\cite{Man69,Vos85,Ram94,Mag97}. They allow one to model many highly
irregular natural objects, which can be viewed as random fractals \cite
{Vos85}. The traces of the Levy motions are also statistically self -
affine, therefore, one may expect that they are also suited for modeling and
studies of natural random fractals.

The stable distributions and the Levy, or Levy - like, random processes are
widely used in different areas, where the phenomena possessing scale
invariance (in a probabilistic sense) are viewed or, at least, can be
suspected, e.g., in economy \cite{Man63,Mgn91,Mgn98}, biology and physiology 
\cite{Wes94}, turbulence \cite{Kla87} and chaotic dynamics \cite{Sch93},
solid state physics \cite{Bou90}, plasma physics \cite{Zim95}, geophysics 
\cite{Sla93} etc. In this respect, the models are needed, which allow one to
simulate fractional Levy motion (fLm) with the prescribed quantitative
statistical properties and to test and improve methods aimed at analysis and
interpretation of experimental data. As far as the authors know, the
properties of fLm are, at least, rarely discussed in physics literature.
Therefore, we believe, that the models constructed ''at the physical level
of strictness'' may be useful for developing effective methods for
experimental data processing and for the purposes of numerical modeling.
When studying this problem, we find that the different levels of complexity
are required for the models aimed to approximate fLm, which differ by their
Levy index $\alpha $ and by their ''long-memory'' behavior, that is,
persistent or anti-persistent one. We discuss this item below and restrict
ourselves with the simplest model, which serves as a good approximation to
the persistent fLm with the Levy index lying between 1 and 2. We propose a
simple approximation to persistent fLm and study its scaling properties.

\section{''$\tau ^{H}$ laws'' for the fractional Levy Motion.}

Our aim is to make an approximation to the process denoted by $L_{\alpha
,H}(t)$, whose increments are stably distributed with the Levy index $\alpha 
$, $0<\alpha \leq 2$, and possess the properties of stationarity and self -
affinity. By analogy with the Definition 3.2 of Ref.\cite{Man68}, the
increments of a random function $L_{\alpha ,H}(t)$ will be said to be self -
affine with parameter $H$ if for any $\kappa >0$ and any $t_{\text{0}}$ 
\begin{equation}
L_{\alpha ,H}(t_0+\kappa \tau )-L_{\alpha ,H}(t_0)\stackrel{d}{=}\left[
\kappa ^H\left( L_{\alpha ,H}(t_0+\tau )-L_{\alpha ,H}(t_0)\right) \right]
\label{1}
\end{equation}

The particular cases are:

(i) ordinary Brownian motion, $L_{\alpha ,H}(t)\equiv B(t)$, which has $%
H=1/2 $;

(ii) fractional Brownian motion, $L_{\alpha ,H}(t)\equiv B_{H}(t)$, which
has $0<H<1$;

(iii) ordinary Levy motion, $L_{\alpha ,H}(t)\equiv L_{\alpha }(t)$, which
has $H=1/\alpha $.

Since the Levy-stable distribution with the Levy index $\alpha $ possesses
the moments of the order $q<\alpha $, the ''$\tau ^H$ laws'' (by terminology
of \cite{Man68}) for the structure functions of the fractional Levy motion
can be stated as follows: for any $0<q<\alpha $%
\begin{equation}
\left\langle \left| L_{\alpha ,H}(t+\tau )-L_{\alpha ,H}(t)\right|
^q\right\rangle ^{1/q}\propto \tau ^H  \label{2}
\end{equation}
where the proportionality coefficient is a function of $q$ and the
parameters of stable distribution for the increments, but not of $H$. The
range of $H$ can be determined as follows. By Minkowski's inequality, for
any positive $\tau _1,\tau _2>0$ and $1<q<\alpha $%
\[
\left\langle \left| L_{\alpha ,H}(t+\tau _1+\tau _2)-L_{\alpha ,H}(t)\right|
^q\right\rangle ^{1/q}\leq 
\]
\[
\left\langle \left| L_{\alpha ,H}(t+\tau _1+\tau _2)-L_{\alpha ,H}(t+\tau
_1)\right| ^q\right\rangle ^{1/q}+ 
\]
\begin{equation}
+\left\langle \left| L_{\alpha ,H}(t+\tau _1)-L_{\alpha ,H}(t)\right|
^q\right\rangle ^{1/q}  \label{3}
\end{equation}
or, using Eq.(2), 
\begin{equation}
(\tau _1+\tau _2)^H\leq \tau _1^H+\tau _2^H  \label{4}
\end{equation}
which implies $H\leq 1$. Further, we require $q$ - th order mean continuity
of $L_{\alpha ,H}(t),$that is \cite{Sko64}, for any $t$ and $q<\alpha $%
\begin{equation}
{\lim_{\Delta t\rightarrow 0}}\left\langle \left| L_{\alpha ,H}(t+\Delta
t)-L_{\alpha ,H}(t)\right| ^q\right\rangle =0  \label{5}
\end{equation}
It implies that $H\geq 0$.

From Eq. (1) the ''$\tau ^H$ law'' follows also for the range of the fLm
with the Levy index $\alpha >1$, 
\begin{equation}
R(\tau )=\sup_{0\leq s\leq \tau }\,(L_{\alpha ,H}(s)-L_{\alpha
,H}(0))-\inf_{0\leq s\leq \tau }(L_{\alpha ,H}(s)-L_{\alpha ,H}(0))
\label{6}
\end{equation}
\begin{equation}
\left\langle R(\tau )\right\rangle \propto \tau ^H  \label{7}
\end{equation}

We make some remarks concerning distinction between the motions with the
Levy index lying between 0 and 1, and those with the index lying between 1
and 2. Comparatively to the latter case, the former one is more involved for
the analysis and numerical simulation. The reasons are as follows:

(i) since $q$ must be less than $\alpha $, the Minkowski's inequality does
not hold for $\alpha \leq 1$, hence we can not use it for establishing the
upper boundary for $H$;

(ii) the statistical mean of the range is infinite for $\alpha \leq 1$;

(iii) with $\alpha $ decreasing, the increments of fLm grow rapidly due to
the power - law tails of stable distributions. Very large increments are
occurring more and more frequently, and the moments of the increments
strongly fluctuate from realization to realization. Thus, the reliability of
the results of simulation decreases. The same conclusion can be done when
treating experimental data, in which the Levy statistics with $\alpha <1$
can, at least, be suspected.

The three peculiarities mentioned above require a modification of the
methods, which we use below for making an approximation to fLm and for
studying its properties. For this reason, we restrict ourselves by the case $%
1\leq \alpha \leq 2$, however, having in mind to study more complicated case 
$0<\alpha <1$ in future. Fortunately for us, there are evidences from
different areas of application, see, e.g., \cite{Man63}\cite{Sla93}, that,
as a rule, the Levy processes with $\alpha >1$ do occur.

\section{The model.}

The process of constructing approximation to fLm can be divided into 3 steps.

Step 1. At the first step we generate random sequence of i.i.d. random
variables possessing stable probability law. These variables play the role
of increments of the ordinary Levy motion having the Levy index $\alpha $, $%
0<\alpha <2$. The value $\alpha =2$ corresponds to the ordinary Brownian
motion, hence in this case the sequence of independent increments is
generated with the use of a standard Gaussian generator. Since in \cite
{Man69} the sequence generated at $\alpha =2$ is called ''approximate
discrete-time white Gaussian noise'' one may call the sequence generated at $%
0<\alpha <2$ ''approximate discrete-time white Levy noise''.

We restrict ourselves by symmetric stable laws with the probability density $%
p_{\alpha ,D}(x)$ and the characteristic function 
\begin{equation}
p_{\alpha ,D}(k)=\left\langle \exp (ikx)\right\rangle =\exp (-D\left|
k\right| ^\alpha )\quad .  \label{8}
\end{equation}
Here $\alpha $ is the Levy index, and $D$ is a positive parameter. At $%
\alpha =1$ and 2 one has the Cauchy and the Gaussian probability laws,
respectively. In other cases the symmetric stable laws are not expressed in
terms of elementary functions. At $0<\alpha <2$ they have power law
asymptotic tails \cite{Gne49}, 
\begin{equation}
p_{\alpha ,D}(x)\propto D\frac{\Gamma (1+\alpha )\sin \left( \pi \alpha
/2\right) }{\pi \left| x\right| ^{1+\alpha }},\quad x\rightarrow \pm \infty
\quad .  \label{9}
\end{equation}
Among the methods of random sequence generation with the given probability
law ${\it F(x)}$ the method of inversion seems most simple and effective 
\cite{Ken39}. However, it is well-known that its validity is limited by the
laws possessing analytic expressions for {\it F}$^{-1}$, hence, the direct
application of the method of inversion to the stable law is not expedient.
In this connection, we exploit an important property of stable
distributions. Namely, such distributions are limiting for those of properly
normalized sums of i.i.d. random variables \cite{Gne49}. To be more
concrete, we generate the needed random sequence in two steps. At the first
one we generate an ''auxiliary'' sequence of i.i.d. random variables $%
\left\{ \xi _j\right\} $, whose distribution density $F^{^{\prime }}(x)$
possesses asymptotics having the same power law dependence as the stable
density with the Levy index $\alpha $ has, see Eq.(9). However, contrary to
the stable law, the function $F(x)$ is chosen as simple as possible in order
to get analytic form of {\it F}$^{-1}$. For example, 
\[
F(x)=\frac 1{2(1+\left| x\right| ^\alpha )},\quad x<0\ \quad , 
\]

\[
F(x)=1-\frac 1{2(1+x^\alpha )}\quad x>0\quad . 
\]
Then, the normalized sum is estimated, 
\begin{equation}
X=\frac 1{am^{1/\alpha }}\sum_{j=1}^m\xi _j\quad ,  \label{10}
\end{equation}
where 
\[
a=\left( \frac \pi {2\Gamma (\alpha )\sin (\pi \alpha /2)}\right) ^{1/\alpha
}\quad . 
\]
According to the Gnedenko theorem on the normal attraction basin to the
stable law \cite{Gne49}, the distribution of the sum (10) converges to the
stable law with the characteristic function (8) and ${\it D}=1$. It is
reasonable to generate the stable sequences with ${\it D}=1$, doing the
rescaling after that, if necessary. Repeating $N$ times the above procedure,
we get a sequence of i.i.d. random variables $X(t),t=1,2,...,N.$ This is an
approximation to a discrete - time white Levy noise.

Step 2. At the second step we convert approximate white Levy noise $X(t)$
into the approximate fractional Levy noise $X_\nu (t)$. For this purpose one
can use the procedure, which is called fractional
integration/differentiation of a white noise \cite{Man69}. We remind the
relation between the Fourier transforms of the function $X(t)$ and of its
fractional integral/derivative $X_\nu (t)$ of the $\nu $-th order: 
\begin{equation}
\stackrel{\wedge }{X}_\nu (\omega )=\frac{\stackrel{\wedge }{X}(\omega )}{%
(-i\omega )^\nu }\quad ,  \label{11}
\end{equation}
where $\nu $ is positive in case of fractional integration and negative in
case of fractional differentiation (to be more accurate, we say about
left-side Riemann - Liouville fractional integral/derivative at the infinite
axis \cite{Sam87}). Fractional integration of a white noise leads to
amplification of low-frequency spectral components of the noise and thus, to
the persistent process, whereas fractional differentiation leads to
amplification of high-frequency spectral components and thus, to the
anti-persistent process. At a qualitative level the difference between the
persistent and anti-persistent behaviors can be formulated as follows: in
the persistent random process the available tendency is supported, whereas
in the anti-persistent process the opposite tendency prevails \cite{Man69}%
\cite{Fed88}. The range of $\nu $ will be discussed below. The approximate
discrete - time fractional Levy noise $X_\nu (t)$ plays the role of the
sequence of increments of the approximation to fLm.

Step 3. With using approximate discrete - time fractional Levy noise $X_\nu
(t)$ the approximation to the fractional Levy motion is defined by 
\begin{equation}
L_{\alpha ,\nu }(t)=\sum_{\tau =1}^tX_\nu (\tau )\quad .  \label{12}
\end{equation}

Let us discuss the restrictions on the possible values of $\nu $. They are
dictated by the restrictions on the range of $H$ for the fLm, which we
intend to approximate.

At first, we note that the approximation to the white Levy noise possesses
the property of self - affinity, 
\begin{equation}
X(\kappa t)\stackrel{d}{=}\kappa ^{1/\alpha }X(t)\quad .  \label{13}
\end{equation}
Here $\kappa $ is a positive integer. It follows from Eq. (13) that 
\begin{equation}
\stackrel{\wedge }{X}%
{\omega  \overwithdelims() \kappa }%
\stackrel{d}{=}\kappa ^{1+1/\alpha }\stackrel{\wedge }{X}(\omega )\quad ,
\label{14}
\end{equation}
and thus, according to our way of modeling, 
\begin{equation}
X_\nu (\kappa t)\stackrel{d}{=}\kappa ^{\nu +1/\alpha }X_\nu (t)\quad .
\label{15}
\end{equation}

This is the property of self - affinity of the increments of the
approximation to fLm. By comparing it with Eq.(1), we conclude that the
equality 
\begin{equation}
H=\nu +1/\alpha  \label{16}
\end{equation}
must hold. Taking into account the restrictions on $H$, we conclude that for 
$1\leq \alpha \leq 2$%
\begin{equation}
-1/\alpha \leq \nu \leq 1-1/\alpha \quad .  \label{17}
\end{equation}

The particular cases of the proposed approximation are as follows:

(i) $\alpha =2,\nu =0.$ In this case we get approximation to the Gaussian
process with independent increments, that is, to the ordinary Brownian
motion, which has the parameter $H$ equal 1/2;

(ii) $\alpha =2,\quad -1/2<\nu <1/2.$ In this case we get approximation to
the fractional Brownian motion, which has the parameter $H$ lying between 0
and 1. We studied the properties of this approximation in \cite{Che02};

(iii) $0<\alpha <2,\quad \nu =0.$ In this case we get approximation to the
ordinary Levy motion, which has the parameter $H$ equal 1/$\alpha .$ The
step 2 is omitted when getting this approximation. Its properties were
studied in \cite{Che01}. The three different models of the ordinary Levy
motions were proposed in \cite{Mai99}. They can be interpreted as a
''difference scheme'' to approximate the evolution equation for the density
distribution of the ordinary Levy motion.

In Fig.1 the admissible range of the index $\nu $ is shown for $1<\alpha
\leq 2$ on the ($\alpha ,\nu )$- plane. This range is bounded by the curve $%
\quad H=\nu +1/\alpha =1$ in the top and by the curve $H=0$ in the bottom.
The right vertical boundary indicated by thick line corresponds to the
fractional Brownian motion, $\alpha =2,-1/2<\nu <1/2.$ The horizontal thick
line $\nu =0$ corresponds to the ordinary Levy motion. This line divides two
regions: upper one, $\nu >0,$ in which the Levy motion is persistent, and
the bottom one, $\nu <0$, in which the motion is anti-persistent. Dotted
lines $a,b$ relate to Figs. 4, 5 and will be explained below.

Before proceeding with the numerical results illustrating the self-affinity
properties of our approximation, we discuss the peculiarities of simulating
persistent and anti-persistent motions. The problem appearing can be
explained by taking fractional Brownian motion as an example \cite{Che02}.
At first we consider the persistent case. Fractional integration of a white
Gaussian noise leads to the fractional Gaussian noise with spectral density
decreasing as frequency increases. Thus, at numerical simulation the
high-frequency inaccuracies are small and the low-frequency ones play the
main role. Now we turn to the anti-persistent case. Fractional
differentiation of a white Gaussian noise leads to the fractional Gaussian
noise with spectral density increasing as frequency increases. At numerical
simulation the high-frequency inaccuracies are prevailing. We have
demonstrated, when studying fractional Brownian motion approximation based
on fractional integration/differentiation of a white noise \cite{Che02},
that the high - frequency inaccuracies are more essential than the low -
frequency ones, that is, the anti - persistent case is modeled with less
accuracy. Moreover, it is known that the other algorithms for simulating
fractional Brownian motion, namely those proposed in \cite{Man69} and in\cite
{Vos85} as well, have the same drawback. Thus, one may expect that the
anti-persistent fLm is also worse modeled. This is indeed so, as we
convinced during numerical simulation. When simulating fractional Brownian
motion, the problem is overcome in part by passing from ''Type -2
Approximation'' to a more sophisticated ''Type 1 Approximation'' having a
finer grid \cite{Man69}. A similar improvement can be performed when doing
fractional differentiation of a white Levy noise. However, such a procedure
requires more analysis, which have to be the subject of a separate paper.
That is why below we present the results for the persistent case only. We
also note that the persistent behavior is prevalent in nature \cite{Man69}%
\cite{Fed88}, thus, we may hope that approximation to the persistent
fractional Levy motion is more needed for applications. Of course, this does
not imply that there is no need for studying anti - persistent fLm. On the
contrary, we have in mind that the complex systems with the feedback require
the development of approximations well-suited for modeling anti - persistent
behavior. As an example we quote long - range anti - correlations and non -
Gaussian behavior of the heartbeat of the healthy subjects \cite{Pen93}.

\section{Numerical results.}

The results of numerical simulation and analysis are shown in Figs. 2 - 5.

In the top of Fig. 2 the probability densities $p(x)$ for the members of the
sequence $X(t)$ are depicted by black points for (a) $\alpha =1.2,$ and (b) $%
\alpha =1.7.$ We use $m=30$ terms in the sum (10). The functions $p_{\alpha
,1}(x)$ obtained by the inverse Fourier transform, see Eq.(8), are shown by
solid lines. In the bottom of Fig. 2 the black points depict asymptotics of
the same probability densities in log - log scale. The solid lines show the
asymptotics given by Eq. (9). It is seen that the Levy index can be
estimated with the use of those values of $X(t)$, which lie outside the peak
located around $x=0$. For the comparison the asymptotics of the Gaussian
distribution with zero mean and unit variance are also shown. The log - log
scale allows one to demonstrate smallness of probability of extreme Gaussian
events. The examples presented demonstrate a good agreement between the
probability densities for the sequences $X(t)$ obtained with the use of the
numerical algorithm proposed and the densities of the stable laws.

We would like to note that the simplicity is a certain merit of the proposed
approximation to a white Levy noise. The approximation is entirely based on
classical formulation of one of the limit theorems and can be easily
generalized for the case of asymmetric stable distributions. It also allows
one, after some modifications, to speed up the convergence to the stable
law. These problems, however, ought to be the subject of a separate paper.
We also note, that two schemes were proposed recently, which use the
combinations of random number generators \cite{Mgn94} and the family of
chaotic dynamical systems with broad probability distributions \cite{Ume98},
respectively.

In Fig. 3 typical samples of approximation to discrete - time Levy noises
with the Levy index $\alpha =1.5$ are depicted for a white noise, $\nu =0$
(at the left), and for a fractional one, $\nu =0.3$ (at the right). In both
cases large frequent ''outliers'' are clearly seen, which are totally absent
in case of the Gaussian noises. An interesting feature of the top figures is
that they look very similar. Indeed, only careful comparison can reveal the
differences between them. However, the corresponding ordinary Levy motion
shown below at the left, and fractional persistent Levy motion shown below
at the right are strikingly different. It is so, because the fractional
noise possesses such correlations, that the values of the noise at different
instants ''acts coherently''. It means that in case of persistent motion,
not only ''jumps'', or ''Levy flights'' (which are clearly visible on both
trajectories) lead to a large departure of the trajectory from the $x$ -
axis, but also ''coordinated action'' of the noise at different instants
results in such a departure. In terms of anomalous diffusion we may argue
that, contrary to the case of the ordinary Levy motion, when an anomalous
diffusion rate is determined almost solely by the ''Levy flights'', in the
persistent Levy motion it is also determined by ''small steps''. This is a
qualitative corollary of the procedure of fractional integration of a white
noise.

In Fig. 4 we illustrate the properties of structure functions. We study $%
\tau $- dependence of the structure functions of our approximation, 
\begin{equation}
\left\langle \left| L_{\alpha ,\nu }(t+\tau )-L_{\alpha ,\nu }(t)\right|
^q\right\rangle ^{1/q}\propto \tau ^s\quad ,  \label{18}
\end{equation}
where $q<\alpha $. Since for the fLm the ''$\tau ^H$ law'' is fulfilled, see
Eq.(1), it is expected, according to our way of modeling, that $s$ is close
to $\nu +1/\alpha $. We demonstrate the results for $q=1/4$. However, we
verified that for any $q$ less than $\alpha $ and not very close to it the
results are just the same. We present the results for the approximation to
the persistent Levy motions, whose parameters $\alpha ,$ $\nu $ vary along
dotted lines $a,b$ shown in Fig. 1.

In Fig. 4a the results for the vertical line $a$ are presented, $\alpha
=1.7. $ The exponent $s$ versus $\nu $, see Eq.(18), is depicted by black
squares. The expected relation $s=\nu +1/\alpha $ is shown by solid line. We
see that numerical results are well fitted by the expected line, but the
discrepancy appears when $\nu $ reaches its upper boundary, that is, for the
strongly persistent case. The same effect appears when simulating fractional
Brownian motion, $\alpha =2$ \cite{Che02}. Since the Fourier transform of a
white noise is divided by $\omega ^\nu ,$ $\nu >0,$ when getting fractional
noise in the persistent case (see Step 2 of the approximation), one may
suppose that the appeared discrepancy is due to the growth of low -
frequency inaccuracies with $\nu $ increasing. However, this is not a single
reason: indeed, we convinced that making longer sample path does not
diminish discrepancy essentially.

In Fig. 4b the results for the processes along the dotted curve $b$ (see
Fig.1) are presented. The curve $b$ corresponds to the case $H=\nu +1/\alpha
=0.8.$ In the figure the $y$- axis indicates the values of $s$, whereas
along the $x$-axis both the values of $\alpha $ (in the bottom, linear
scale) and of $\nu $ (in the top, non - linear scale) are shown. Black
squares indicate the values of $s$ measured along the curve $b$, whereas
solid line indicates the value $s=0.8,$ which we expect to get. We see, that
numerical results are well fitted by the expected line. For the comparison
we show by crosses the exponent $s$ measured from Eq. (18) for the second
order structure function, $q=2.$ The theoretical second order structure
function is infinite for the Levy motion. This circumstance has two
consequences for the numerical simulation and/ or at experimental data
processing. The first one is that the second order structure function
increases with the length of the Levy motion trajectory increasing. The
second consequence is that the exponent $s$ does not depend on $\alpha $ and
demonstrates ''pseudo - Gaussian'' behavior. Indeed, the dotted line $s=\nu
+1/2$ (which indicates the relation between $s$ and $\nu $ for the
approximation to the fractional Brownian motion) well fits the relation
indicated by crosses in Fig. 4b.

In Fig.5 we illustrate the studies of the range. We investigate $\tau $ -
dependence of the range $R_\nu $ of our approximation to the fLm with the
Levy index $\alpha >1,$%
\begin{equation}
R_\nu (\tau )=\sup_{0\leq s\leq \tau }\left[ L_{\alpha ,\nu }(s)-L_{\alpha
,\nu }(0)\right] -\inf_{0\leq s\leq \tau }\left[ L_{\alpha ,\nu
}(s)-L_{\alpha ,\nu }(0)\right] \quad ,  \label{19}
\end{equation}
\begin{equation}
\left\langle R_\nu (\tau )\right\rangle \propto \tau ^{H_\nu }\quad .
\label{20}
\end{equation}
Since the ''$\tau ^H$ law'' is fulfilled for the range of the Levy motion,
see Eqs.(6,7), it is expected, according our way of modeling, that $H_\nu $
is close to $H=\nu +1/\alpha .$

In the empirical rescaled range analysis, that is, at experimental data
processing, or in numerical simulation, the range of the random process is
divided by the standard deviation of its increments after subtraction of a
linear trend \cite{Hur65}. This procedure, called the Hurst method, or the
method of normalized range, in particular, smooths the variations of the
range on different segments of time series. As the result of the empirical
rescaled range analysis of experimental data, one gets the Hurst exponent of
the process, that is, the exponent $H_\nu $ in our case. However, the Hurst
method is not satisfactory for the Levy motion because of the infinity of
the theoretical value of the standard deviation. Therefore, we propose to
modify the Hurst method by exploiting the $1/\alpha $ - th root of the $%
\alpha $ - th moment instead of standard deviation, that is, 
\begin{equation}
\sigma _\alpha =\left( \frac 1\tau \sum_{t=1}^\tau \left| X(t)\right|
^\alpha \right) ^{1/\alpha }  \label{21}
\end{equation}
Since it has only weak logarithmic divergence with the number of terms in
the sum increasing, then the power - law dependence (20) is not changed.

In Fig. 5a we present the results of applying the modified Hurst method to
the ranges of the processes with the parameters along the vertical line $a$
in Fig. 1. The Levy index $\alpha $ is equal 1.7. In the figure the exponent 
$H_\nu $ versus $\nu $ is depicted by black squares. The relation $H_\nu
=\nu +1/\alpha $ is shown by solid line. As in case of the structure
function, see Fig. 4a, we notice that numerical results are well fitted by
the line showing the expected relation. However, the discrepancy appear for $%
\nu $ reaching its upper boundary, that is, for the strongly persistent
case. In this respect we may repeat our above comment to the analogous
discrepancies in case of the structure function.

In Fig. 5b we present the results of applying the modified Hurst method to
the ranges of the process with the parameters changing along the curve $b$
in Fig. 1. This curve corresponds to the case $H=\nu +1/\alpha =0.8.$ The $y$
- axis indicates the values of $H_\nu $ , whereas along the $x$ - axis both
the values of $\alpha $ (in the bottom, linear scale) and of $\nu $ (in the
top, non - linear scale) are shown. Black squares indicate the values of $%
H_\nu $ measured along the curve $b$, whereas solid line indicates the value 
$H_\nu =0.8$, which we expect to get. We see the discrepancy between the
numerical results and the expected line. However, the discrepancy is even
larger, if one uses the ordinary, or non - modified, Hurst method. The
results of its exploiting are shown by crosses, whereas the ''pseudo -
Gaussian'' relation $H_\nu =\nu +1/2$ (which does not depend on $\alpha $)
is depicted by dotted line. One can clearly see the inapplicability of the
''traditional'' Hurst method for characterizing properties of the Levy
motions.

\section{Results}

The results of the paper are as follows.

1. We propose a model for the random process, whose increments are
stationary, self - affine and distributed with the stable probability law.
By analogy with the fractional Brownian motion, the family of these
processes can be called fractional Levy motion.

2. When constructing our model, the two basic steps are:

(i) the use of the Gnedenko limit theorem for the normal attraction basin of
a stable law; the theorem gives us a simple way for generating a sequence of
independent stably distributed variables, which approximate a discrete -
time white Levy noise;

(ii) fractional integration/differentiation of a white noise; this procedure
converts white noise into a fractional one, thus allowing us to approximate
a discrete - time fractional Levy noise.

3. We find the ranges of the change of the order of fractional
integration/differentiation, inside of which the increments of the Levy
motion approximation obtained possess the property of self - affinity. This
property manifests itself in so - called ''$\tau ^{H}$ laws'', that is, in
the power - law time - dependence of both the structure function of the
order $q$ less than the Levy index $\alpha $ and the range. The relation
between the exponent $H$ and the order of fractional
integration/differentiation is also obtained.

4. We study the ''$\tau ^{H}$ law'' for the structure functions of the
approximation and find a good agreement between numerical results and the
theory for the $q$ - th order structure function, $q<\alpha .$ We also
demonstrate that the second order structure function has a ''pseudo -
Gaussian'' time behavior, which is irrespective of the Levy index $\alpha .$

5. We study the ''$\tau ^{H}$ law'' for the range and demonstrate, that the
''traditional'' Hurst method exploiting the normalized rescaled range leads
to a ''pseudo - Gaussian'' time - dependence, which is irrespective of the
Levy index $\alpha .$ Thus, we propose the modified Hurst method, in which
the $1/\alpha $ - th root of the $\alpha $ - th moment of the sequence of
increments is used instead of standard deviation. The modified Hurst method
leads to the rescaled range time - dependence, which is much closer to the
theory than that of normalized range.

6. We conclude that our approximation is suitable for simulating persistent
Levy motion with the Levy indexes varying between 1 and 2. We also discuss
the possible reasons why our model works worse in the anti - persistent
region and in the region of Levy indexes less than 1.

7. As it concerns with the ''pseudo - Gaussian'' effects described above,
they allow us to suggest that at estimating the second order structure
function and normalized span from experimental data the ''Levy nature '' of
them can be easily masked. This, in turn, rises an interesting task of
developing statistical methods for extracting reliable characteristics from
experimental data, for which Levy statistics can be expected from, e.g.,
some physical reasons.

\acknowledgments

This work was supported in finance by National Academy of Science of
Ukraine, the Project ``Chaos-2`` and by INTAS Program, the Projects 93-1194,
LA-96-09 and 98 - 01.

\newpage

Fig. 1. Shaded region indicates the admissible range of the order $\nu \,$of
fractional integration/differentiation of a white Levy noise. The Levy index 
$\alpha $ is varied from 1 to 2. The vertical thick line at the right
corresponds to the fractional Brownian motion, $\alpha =2,$ $-1/2<\nu <1/2.$
The horizontal thick line $\nu =0$ corresponds to the ordinary Levy motion.
Above the horizontal line the process is persistent, whereas below it is
anti - persistent. Along the vertical dotted line $a$ the Levy index $\alpha 
$ is 1.7. Along the dotted line $b$ the relation $\nu +1/\alpha =0.8$ holds.
The self - similarity properties of the approximation to fLm along the
dotted lines are illustrated in Figs. 4, 5.

Fig. 2. Probability densities (above) and their asymptotics (below) are
shown for the approximation to discrete - time white Levy noises with the
Levy indexes $\alpha =1.2$ and $\alpha =1.7.$ The probability densities and
the asymptotics of the stable laws are indicated by solid lines. Below for
the qualitative comparison the asymptotics of the Gaussian distribution are
shown.

Fig. 3. Above: typical samples of approximation to a discrete - time white
Levy noise, $\nu =0$ (at the left) and fractional Levy noise, $\nu =0.3 $
(at the right). Below: corresponding trajectories of approximation to the
ordinary Levy motion (at the left) and the fractional Levy motion (at the
right).

Fig. 4. (a) Exponent $s$ of the structure function of order $q=1/4$ versus
the order of fractional integration $\nu $, see Eq.(18) (black points).
Processes along the vertical dotted line $a$ from Fig.1 are investigated.
The Levy index $\alpha $ is 1.7. Solid line indicates the relation $s=\nu
+1/\alpha .$ (b) Exponent $s$ is indicated for the processes along the
dotted line $b$ from Fig.1 (black points). The parameters $\alpha $ , $\nu $
of these processes obey the law $\nu +1/\alpha =0.8.$ The order $q$ of the
structure function is 1/4. Solid line indicates the relation $s=0.8.$ The
exponent $s$ for the second order structure function is indicated by
crosses. Dotted line shows the ''pseudo - Gaussian'' relation $s=\nu +1/2.$

Fig. 5. (a) Exponent $H_{\nu }$ for the range, see Eqs. (19), (20), versus $%
\nu $ is depicted by black points. The numerical results are obtained with
the use of the modified Hurst method proposed in the paper. Processes along
the vertical dotted line $a$ from Fig.1 are investigated. The Levy index $%
\alpha $ is 1.7. Solid line indicates the relation $H_{\nu }=\nu +1/\alpha .$
(b) Exponent $H_{\nu }$ is indicated for the processes along the dotted line 
$b$ from Fig.1 (black points). The parameters of these processes obey the
law \strut $\nu +1/\alpha =0.8.$ The modified Hurst method is used. Solid
line indicates the relation $H_{\nu }=0.8.$ The exponent $H_{\nu }$,
obtained with the use of ''traditional'' Hurst method, is indicated by
crosses. Dotted line shows the ''pseudo - Gaussian'' relation $s=\nu +1/2.$

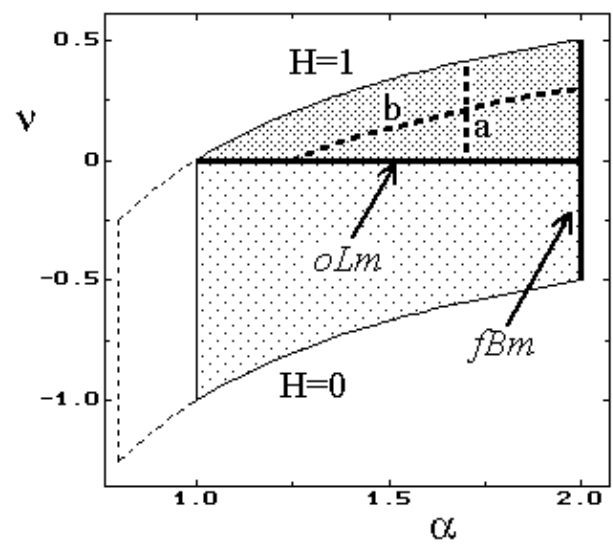
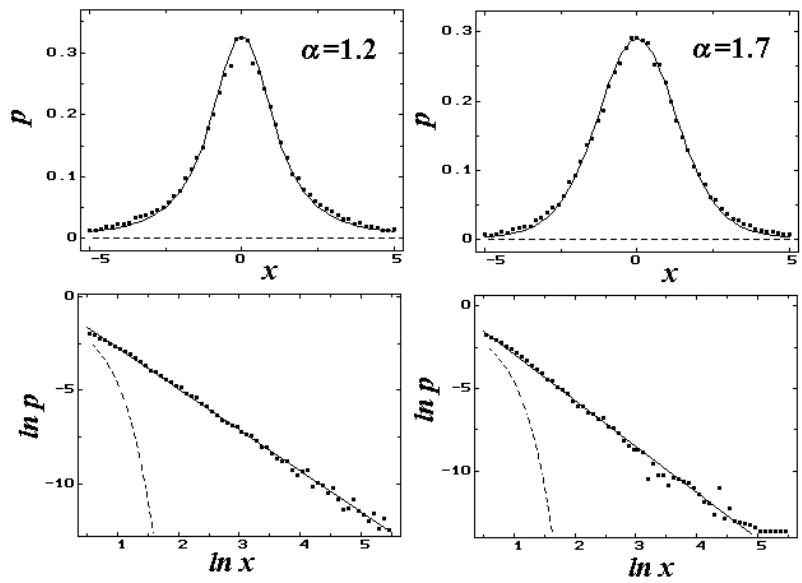
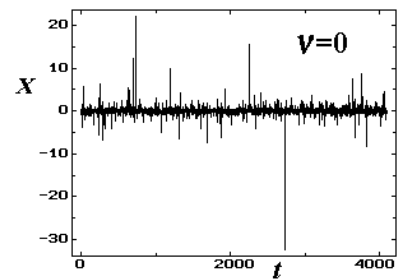
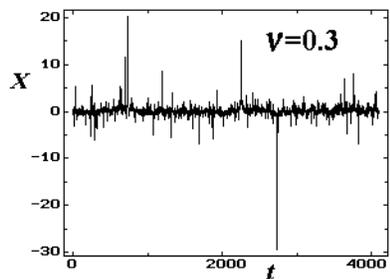
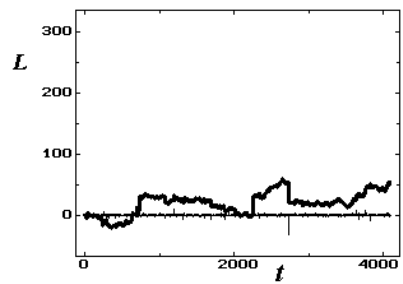
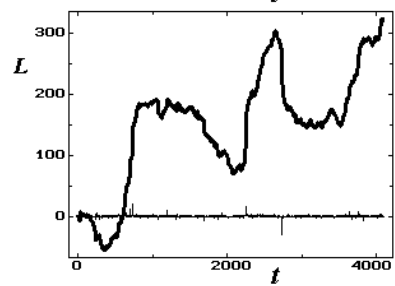
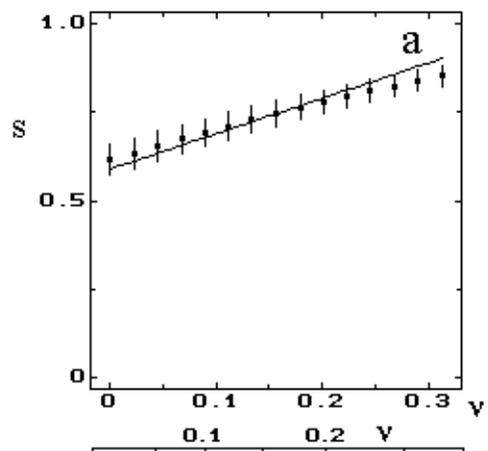
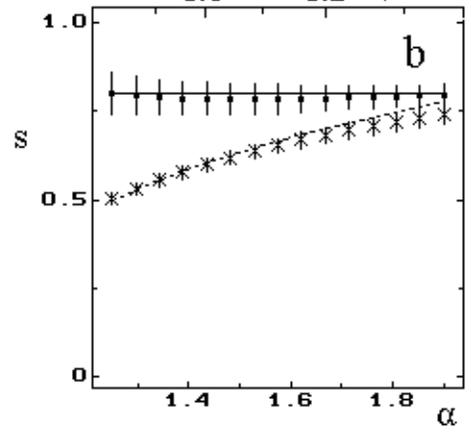
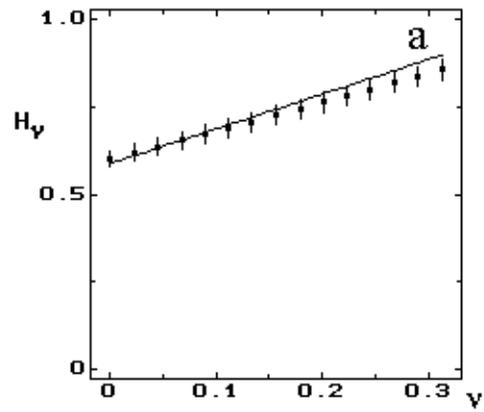
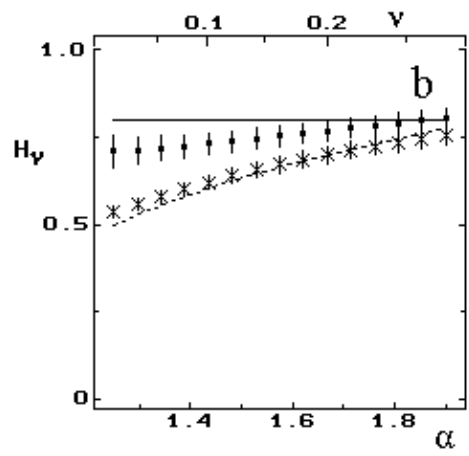
\end{document}